\documentclass[twocolumn]{aastex631}

\usepackage[utf8]{inputenc}

\usepackage{amsmath}
\usepackage{array}   
\usepackage{enumitem}
\usepackage{CJK}
\usepackage{hyperref}
\received{}
\revised{}
\accepted{}
\submitjournal{\apj}

\shorttitle{MIRI Imaging of VV 114}
\shortauthors{Evans et al.}

\begin{document}

\title{GOALS-JWST: Hidden Star Formation and Extended PAH Emission in the Luminous Infrared Galaxy VV 114}

\correspondingauthor{A. S. Evans}
\email{aevans@virginia.edu}

\author[0000-0003-2638-1334]{A. S. Evans}
\affil{Department of Astronomy, University of Virginia, 530 McCormick Road, Charlottesville, VA 22903, USA}
\affiliation{National Radio Astronomy Observatory, 520 Edgemont Road, Charlottesville, VA 22903, USA}

\author[0000-0003-1924-1122]{D. T. Frayer}
\affiliation{Green Bank Observatory, 155 Observatory Road, Green Bank, WV 24944, USA}

\author[0000-0002-2688-1956]{Vassilis Charmandaris}
\affiliation{Department of Physics, University of Crete, Heraklion, 71003, Greece}
\affiliation{Institute of Astrophysics, Foundation for Research and Technology-Hellas (FORTH), Heraklion, 70013, Greece}
\affiliation{School of Sciences, European University Cyprus, Diogenes street, Engomi, 1516 Nicosia, Cyprus}

\author[0000-0003-3498-2973]{Lee Armus}
\affiliation{IPAC, California Institute of Technology, 1200 E. California Blvd., Pasadena, CA 91125}

\author[0000-0003-4268-0393]{Hanae Inami}
\affiliation{Hiroshima Astrophysical Science Center, Hiroshima University, 1-3-1 Kagamiyama, Higashi-Hiroshima, Hiroshima 739-8526, Japan}

\author[0000-0001-7291-0087]{Jason Surace}
\affiliation{IPAC, California Institute of Technology, 1200 E. California Blvd., Pasadena, CA 91125}

\author[0000-0002-1000-6081]{Sean Linden}
\affiliation{Department of Astronomy, University of Massachusetts at Amherst, Amherst, MA 01003, USA}

\author{B. T. Soifer}
\affil{Spitzer Science Center, California Institute of Technology, 1200 E. California Blvd., Pasadena, CA 91125 USA}

\author[0000-0003-0699-6083]{Tanio Diaz-Santos}
\affiliation{Institute of Astrophysics, Foundation for Research and Technology-Hellas (FORTH), Heraklion, 70013, Greece}
\affiliation{School of Sciences, European University Cyprus, Diogenes street, Engomi, 1516 Nicosia, Cyprus}

\author[0000-0003-3917-6460]{Kirsten L. Larson}
\affiliation{AURA for the European Space Agency (ESA), Space Telescope Science Institute, 3700 San Martin Drive, Baltimore, MD 21218, USA}

\author[0000-0002-5807-5078]{Jeffrey A. Rich}
\affiliation{The Observatories of the Carnegie Institution for Science, 813 Santa Barbara Street, Pasadena, CA 91101}

\author[0000-0002-3139-3041]{Yiqing Song}
\affiliation{Department of Astronomy, University of Virginia, 530 McCormick Road, Charlottesville, VA 22903, USA}
\affiliation{National Radio Astronomy Observatory, 520 Edgemont Road, Charlottesville, VA 22903, USA}

\author[0000-0003-0057-8892]{Loreto Barcos-Munoz}
\affiliation{National Radio Astronomy Observatory, 520 Edgemont Road, Charlottesville, VA 22903, USA}
\affiliation{Department of Astronomy, University of Virginia, 530 McCormick Road, Charlottesville, VA 22903, USA}

\author[0000-0002-8204-8619]{Joseph M. Mazzarella}
\affiliation{IPAC, California Institute of Technology, 1200 E. California Blvd., Pasadena, CA 91125}

\author[0000-0003-3474-1125]{George C. Privon}
\affiliation{National Radio Astronomy Observatory, 520 Edgemont Road, Charlottesville, VA 22903, USA}
\affiliation{Department of Astronomy, University of Florida, P.O. Box 112055, Gainesville, FL 32611, USA}

\author[0000-0002-1912-0024]{Vivian U}
\affiliation{Department of Physics and Astronomy, 4129 Frederick Reines Hall, University of California, Irvine, CA 92697, USA}

\author[0000-0001-7421-2944]{Anne M. Medling}
\affiliation{Department of Physics \& Astronomy and Ritter Astrophysical Research Center, University of Toledo, Toledo, OH 43606, USA}
\affiliation{ARC Centre of Excellence for All Sky Astrophysics in 3 Dimensions (ASTRO 3D)}

\author[0000-0002-5666-7782]{Torsten B\"oker}
\affiliation{European Space Agency, Space Telescope Science Institute, Baltimore, Maryland, USA}

\author[0000-0002-5828-7660]{Susanne Aalto}
\affiliation{Department of Space, Earth and Environment, Chalmers University of Technology, 412 96 Gothenburg, Sweden}

\author[0000-0002-4923-3281]{Kazushi Iwasawa}
\affiliation{Institut de Ci\`encies del Cosmos (ICCUB), Universitat de Barcelona (IEEC-UB), Mart\'i i Franqu\`es, 1, 08028 Barcelona, Spain}
\affiliation{ICREA, Pg. Llu\'is Companys 23, 08010 Barcelona, Spain}

\author[0000-0001-6028-8059]{Justin H. Howell}
\affiliation{IPAC, California Institute of Technology, 1200 E. California Blvd., Pasadena, CA 91125}

\author{Paul van der Werf}
\affiliation{Leiden Observatory, Leiden University, NL-2300 RA Leiden, Netherlands}

\author{Philip Appleton}
\affiliation{IPAC, California Institute of Technology, 1200 E. California Blvd., Pasadena, CA 91125}

\author{Thomas Bohn}
\affiliation{Hiroshima Astrophysical Science Center, Hiroshima University, 1-3-1 Kagamiyama,  Higashi-Hiroshima, Hiroshima 739-8526, Japan}

\author[0000-0002-1207-9137]{Michael J. I. Brown}
\affiliation{School of Physics \& Astronomy, Monash University, Clayton, VIC 3800, Australia}

\author[0000-0003-4073-3236]{Christopher C. Hayward}
\affiliation{Center for Computational Astrophysics, Flatiron Institute, 162 Fifth Avenue, New York, NY 10010, USA}

\author{Shunshi Hoshioka}
\affiliation{Hiroshima Astrophysical Science Center, Hiroshima University, 1-3-1 Kagamiyama,  Higashi-Hiroshima, Hiroshima 739-8526, Japan}

\author{Francisca Kemper}
\affiliation{ICREA, Pg. Llu\'is Companys 23, 08010 Barcelona, Spain}

\author[0000-0001-8490-6632]{Thomas Lai}
\affiliation{IPAC, California Institute of Technology, 1200 E. California Blvd., Pasadena, CA 91125}

\author{David Law}
\affiliation{Space Telescope Science Institute, 3700 San Martin Drive, Baltimore, MD, 21218, USA}

\author[0000-0001-6919-1237]{Matthew A. Malkan}
\affiliation{Department of Physics \& Astronomy, UCLA, Los Angeles, CA 90095-1547}

\author{Jason Marshall}
\affiliation{Glendale Community College, 1500 N. Verdugo Rd., Glendale, CA 91208}

\author{Eric J. Murphy}
\affiliation{National Radio Astronomy Observatory, 520 Edgemont Road, Charlottesville, VA 22903, USA}

\author{David Sanders}
\affiliation{Institute for Astronomy, University of Hawaii, 2680 Woodlawn Drive, Honolulu, HI 96822}

\author{Sabrina Stierwalt}
\affiliation{Occidental College, Physics Department, 1600 Campus Road, Los Angeles, CA 90042}


\begin{abstract}
{\it James Webb Space Telescope} ({\it JWST}) Mid-InfraRed Instrument (MIRI) images of the luminous infrared (IR) galaxy VV 114 are presented. This redshift $\sim 0.020$ merger has a western component (VV 114W) rich in optical star clusters and an eastern component (VV 114E) hosting a luminous mid-IR nucleus hidden at UV and optical wavelengths by dust lanes. With MIRI, the VV 114E nucleus resolves primarily into bright NE and SW cores separated by 630 pc.
This nucleus comprises 45\% of the 15$\mu$m light of VV 114, with the NE and SW cores having IR luminosities, $L_{\rm IR} (8-1000\mu{\rm m}) \sim 8\pm0.8\times10^{10}$ L$_\odot$ and $\sim 5\pm0.5\times10^{10}$ L$_\odot$, respectively, and IR densities, $\Sigma _{\rm IR} \gtrsim 2\pm0.2\times10^{13}$ L$_\odot$ kpc$^{-2}$ and $\gtrsim 7\pm0.7\times10^{12}$ L$_\odot$ kpc$^{-2}$, respectively -- in the 
range of $\Sigma _{\rm IR}$ for the Orion star-forming core
and the nuclei of Arp 220.  
The NE core, previously speculated to have an Active
Galactic Nucleus (AGN), has starburst-like mid-IR colors. In contrast, the VV 114E SW has AGN-like colors.
Approximately 40 star-forming knots
with $L_{\rm IR} \sim 0.02-5\times10^{10}$ L$_\odot$
are identified, 25\% of which
have no optical counterpart. 
Finally, diffuse emission accounts for 40--60\%
of the mid-IR emission. Mostly notably, filamentary Poly-cyclic Aromatic Hydrocarbon (PAH) emission stochastically excited by UV and optical photons accounts for half of the 7.7$\mu$m light of VV 114.
This study illustrates the ability of 
{\it JWST} to detect obscured compact activity and distributed PAH emission in
the most extreme starburst galaxies in the local Universe.
\end{abstract}

\keywords{Active galaxes, Galaxy nuclei, Luminous infrared galaxies, Star forming regions}

\section{Introduction}

The $L_{\rm IR} [8-1000\mu{\rm m}] = 4.5\times10^{11}$ L$_\odot$ luminous infrared galaxy (LIRG) VV 114 (= IC 1623 = Arp 236) is a prime target for Early Science observations with {\it JWST}. It is a striking example of 
how dust obscuration can hide the true nature of a luminous galaxy and cause a drastic transformation in appearance from the UV through the IR.
VV 114 is a mid-stage merger in the Great Observatories all Sky LIRG Survey \citep[GOALS:][]{armus2009} consisting of two nuclei separated by
$\sim 20\arcsec$ (8 kpc).
The western component of VV 114 (VV 114W) is bright in UV and optical
light \citep[][Figures \ref{fig:figcolor}a and \ref{fig:figgray}a]{knop1994,goldader2002} and hosts a large number of optically-luminous young star clusters \citep{linden2021}. 
The eastern component, VV 114E, is invisible at UV wavelengths and has prominent dust lanes which cover much of the diffuse light of the underlying stellar population at optical wavelengths. Beyond 1$\mu$m, VV 114E increasingly becomes the dominant luminosity component of the merger \citep[][]{knop1994,doyon1995,scoville2000,soifer2001,lefloch2002}. Indeed, the total IR-to-UV luminosity ratio of VV 114 is 11.2 \citet[][]{howell2010}, indicating that VV 114E generates most of the energy in the merger. More specifically, the 7$\mu$m-to-UV flux density ratio of VV 114E is $\sim 800$, with the corresponding ratio of VV 114W being just $\sim 10$ \citep{charmandaris2004}. This color contrast of the two components makes VV 114 the most extreme case ever observed among local LIRGs, and may have important implications on the interpretation of colors of high-z IR luminous interacting systems \citet[]{charmandaris2004}. 
While the most sensitive mid-IR images of VV 114 to date have been obtained with space-based telescopes ({\it ISO}, {\it Spitzer}: Figure \ref{fig:figgray}b,c) at $2-6\arcsec$ resolutions, high-resolution ($0.34\arcsec$) Keck MIRLIN observations detected only the VV 114E nucleus at 12.5$\mu$m due to limited sensitivity are these wavelengths from the ground, but were able to resolve it into a NE and SW core separated by $\sim 2\arcsec$ \citep[$\sim 800$ pc;][]{soifer2001}. Radio images of VV 114 are further resolve the SW core into four sources at the
highest resolution \citep[][]{condon1991,soifer2001,song2022}. The NE and SW cores of VV 114E are detected in the rotational transition of CO, with a significant fraction of the cold molecular gas (traced by sub/millimeter-wave rotational molecular transitions) and cold dust far-IR emission emanate from
the overlap region between the two galaxies \citep[e.g.,][]{yun1994,frayer1999,saito2018}.
Finally, the power source of the heavily obscured nucleus in VV 114E has been difficult to
diagnose, with the primary evidence of an energetically important AGN coming
from mid-IR colors \citep{lefloch2002} and the hard X-ray spectrum \citep{grimes2006} of VV 114E (see also Rich et al, in prep).

In this {\it Letter}, {\it JWST} Early Release Science 
imaging observations of VV 114 with the 
Mid InfraRed Instrument (MIRI) are presented. With a 
spatial resolution of $0.18-0.48\arcsec$ over 
the wavelength range $5.6-15\mu$m, {\it JWST} (6.5m-aperture) 
provides a factor of $\sim 8$ higher resolution than {\it Spitzer} (0.85m-aperture) 
and nearly two orders of magnitude in sensitivity, 
enabling the highest resolution and most sensitive observations to date of the
energy sources that collectively power the high IR luminosity 
of VV 114.
A cosmology with $H_0 = 70$ km s$^{-1}$ Mpc$^{-1}$, $\Omega_\Lambda = 0.72$, and 
$\Omega_{\rm matter} = 0.28$ is adopted. At the redshift of VV 114 ($z=0.020$), $1\arcsec$ subtends 400 pc.

\section{Observations and Reduction}
Observations of VV 114 were obtained with the MIRI instrument \citep{rieke2015, bouchet2015} on {\it JWST} on 2022 July 02 (Program ID 1328; PI Armus). Observations were taken with the F560W ($\lambda _0 = 5.6\mu$m), F770W ($\lambda _0 = 7.7\mu$m) and F1500W ($\lambda _0 = 15\mu$m) filters using both the ``BRIGHTSKY'' Subarray mode ($56.3\arcsec \times 56.3\arcsec$ = $512\times512$ pixels) centered to image V114E and VV 114W and the SUB128 Subarray mode (SUB128: $14.1\arcsec \times 14.1\arcsec$ =  $128\times136$ pixels) centered at the peak of the mid-IR emission in VV 114E.  
The SUB128 Subarray mode observations use shorter exposures designed to avoid saturation from the brightest regions.
A three-point dither pattern was used for each Subarray mode per filter, and the total exposure times in F560W, F770W and F1500W with the BRIGHTSKY Subarray mode are
348s, 114s and 114s, respectively, and are 46s, 48s and 48s, respectively, for the SUB128 Subarray mode.
The images presented here are Level 3 products produced through the standard {\it JWST} Science Calibration Pipeline using the software version (CRDS\_VER) 11.16.3.

\section{Photometry}

In order to derive photometry for sources in the MIRI 
images, 
the python routine FIND was 
run with a $25\sigma$ threshold and FHWM of 4.36 pixels 
on the F1500W MIRI image to identify unresolved sources.
Photometry was then measured for the sources in each image in 5-pixel ($= 0.55\arcsec$)  radius apertures and 7-9 pixel ($= 0.77-0.99\arcsec$) radius annuli for local
background subtraction using the IDL routine APER. For the NE and SW
cores of VV 114E, an 8-pixel ($= 0.88\arcsec$) radius aperture and a 9-11 
($= 0.99-1.1\arcsec$) pixel radius
annuli for local background subtraction were adopted. 
Uncertainties were measured by
varying the radius of the annuli over which the background estimates were measured.
Aperture corrections were calculated using
the encircled aperture-to-total energy for synthetic point spread functions provided
on the STScI website\footnote{\url{https://jwst-docs.stsci.edu}}. The resultant photometric estimates are provided in Table \ref{tab:phot}. For measurements
of the total emission in VV 114, the IPAC package 
SKYVIEW v3.8 was utilized.
Measurements were done with a circular ($0.4378\arcmin$-radius) with background subtraction.

\begin{figure*}[htb]
    \centering
    \includegraphics[scale=0.17]{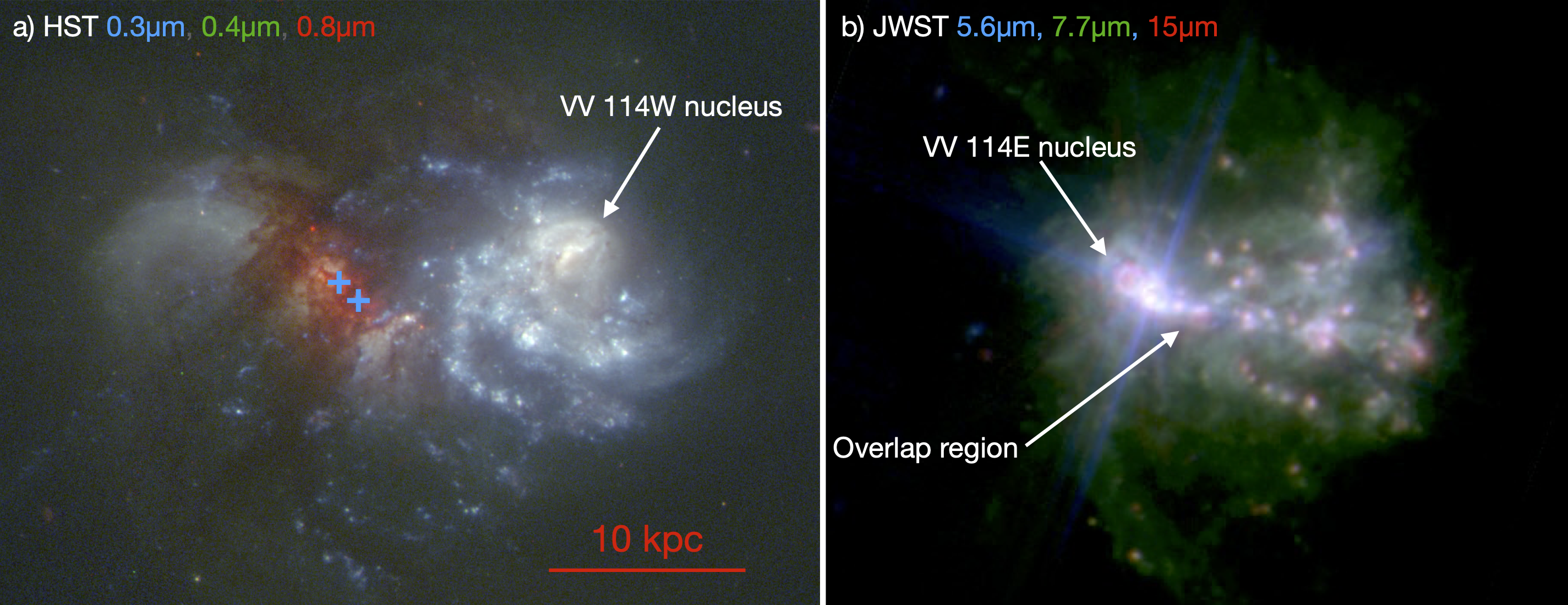}
    \caption{(a) False-color optical image of VV 114 constructed with {\it HST} WFC3/UVIS F330W (0.3$\mu$m) and the ACS/WFC F435W (0.4$\mu$m) and F814W (0.9$\mu$m)
    data. Both images have the same spatial scale. The blue ``+'' symbols mark the location of the VV 114E NE and SW cores detected with MIRI. (b) False color mid-IR image constructed with F560W, F770W and F1500W MIRI data. Blue (5.6$\mu$m) diffraction spikes emanate from the VV114E NE and SW cores. The images are displayed in a logarithmic stretch to highlight both high and low surface brightness emission, and they show the optical star formation
    (a - white knots) and dust lanes (a - red), the reddened nucleus and star-forming regions (b - pink), and the extended, filamentary 7.7$\mu$m PAH emission (b - green) which covers much of the MIRI field of view.\label{fig:figcolor}}
\end{figure*}

\begin{figure*}[htb]
    \centering
    \includegraphics[scale=0.22]{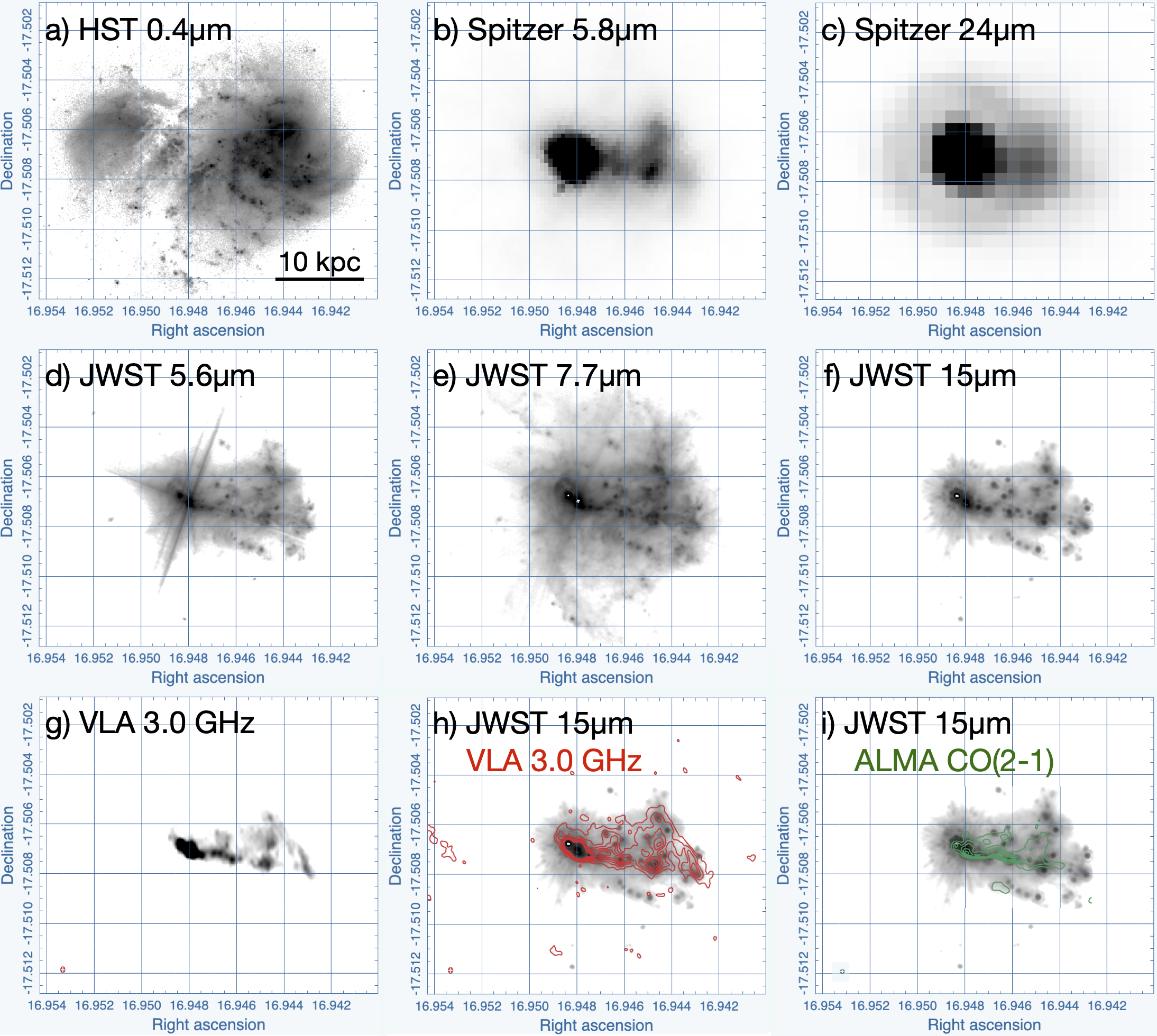}
    \caption{A comparison of the {\it JWST} MIRI 5.6$\mu$m, 7.7$\mu$m and 15$\mu$m images to the {\it HST} 0.4$\mu$m, {\it Spitzer} IRAC 5.8$\mu$m and MIPS 24$\mu$m, VLA 3.0 GHz ($\theta_{\rm FWHM} \sim 0.7\arcsec$ resolution), and ALMA CO(2-1) emission ($\theta_{\rm FWHM} \sim 0.5\arcsec$: green contours in i). The axes are in degrees of Right Ascension and Declination. The {\it HST} and {\it JWST} images are displayed in a logarithmic stretch to highlight both high and low surface brightness emission. The {\it Spitzer} and VLA images are displayed in a linear stretch. The eastern dust lanes visible in the 0.4$\mu$m image cover the bright mid-infrared emission observed with {\it Spitzer} and {\it JWST}. Both VV 114E cores are also visible in the VLA images, as well as several of the higher surface brightness star-forming regions observed in the {\it JWST} images (i). The {\it JWST} images show an order of magnitude improvement in resolution and a factor of 50 improvement in sensitivity relative to {\it Spitzer}, allowing for the identification of $\sim 40$ mid-IR star-forming regions and underlying diffuse emission at resolutions of $\sim 100-200$ pc. \label{fig:figgray}}
\end{figure*}

\section{Results}

\subsection{VV 114E Nucleus and Star-Forming Regions \label{sec:knots}}

\begin{figure}[htb!]
    \centering
    \includegraphics[scale=0.09]{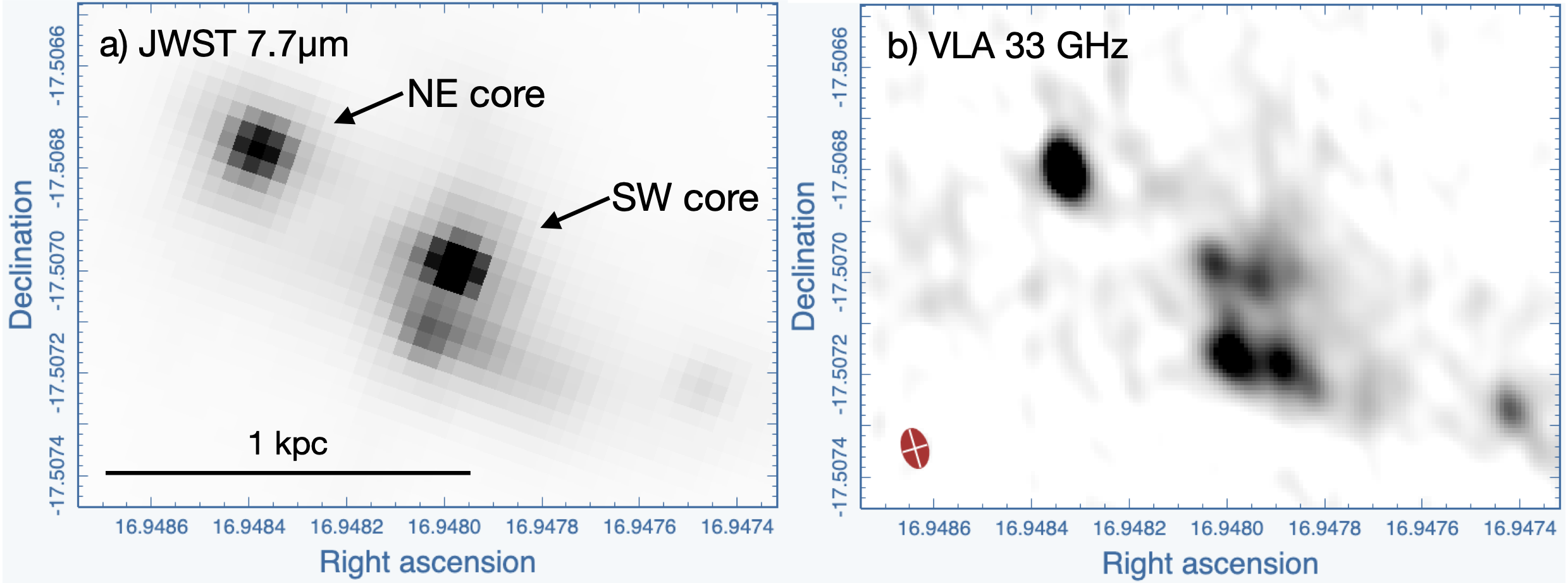}
    \caption{A close-up comparison of the VV 114E nucleus at 7.7$\mu$m and 33 GHz ($\theta_{\rm FWHM} \sim 0.2\arcsec$: \citet{song2022}). The NE core and the western-most knot in the lower right-hand side of the MIRI images are offset from their VLA counterparts by $0.14\arcsec$. In the {\it JWST} image, the nucleus is separated into a NE and SW core. The SW core is further resolved into two components, with the most southern component having a tail of emission extending in the SW direction.
    In the VLA image, the SW core is resolved into four components.\label{fig:fignuc}}
\end{figure}

Figure \ref{fig:figcolor} is a comparison of false-color {\it HST} optical 
and {\it JWST} 
images of VV 114.
Figure \ref{fig:figgray} is a comparison of the MIRI images to the {\it HST} 0.4$\mu$m image \citep[][]{Kim2013}, 
archival {\it Spitzer} IRAC CH3
(5.8$\mu$m), {\it Spitzer} MIPS CH1 (24$\mu$m), ALMA
CO(2-1) and VLA 3.0 GHz images \citep[][]{song2022}. The region of VV 114E covered by dust lanes in the optical image (Figure \ref{fig:figcolor}a and \ref{fig:figgray}a) contains the brightest mid-IR sources. In the MIRI images (Figure \ref{fig:figcolor}b, \ref{fig:figgray}d,e,f and \ref{fig:fignuc}a), the unresolved bright mid-IR nucleus in VV 114E is resolved into two primary cores (which are saturated in the BRIGHTSKY mode images) separated by 630 pc (= $1.6\arcsec$), consistent with the Keck 12.5$\mu$m MIRLIN data \citep{soifer2001}. The SW core is further resolved into a N and S component with a separation of 0.45$\arcsec$ (= 180 pc: Figure \ref{fig:fignuc}a), and the S component has a tail of emission that extends to the SW. (For the photometric results, the S component is treated as part of the SW core.) 
In the VLA 33 GHz image, the SW core is resolved into 4 components, and thus the tail of emission in the 7.7$\mu$m image corresponds to a separate component at 33 GHz (e.g., Figure \ref{fig:fignuc}b). The measured flux density ratios of the NE to SW core are
$\sim 0.3$, 0.6 and 1.0 at 5.6$\mu$m, 7.7$\mu$m and 15$\mu$m, respectively. 


The extended mid-IR emission observed with IRAC and MIPS (Figure \ref{fig:figgray}b,c) west
of the VV 114E nucleus is partially resolved by MIRI into $\sim 40$ compact knots. These knots collectively comprise $\sim 20$\% of the light at 15$\mu$m.
A quarter of the knots
have no optical {\it HST} 0.4$\mu$m and/or 0.9$\mu$m counterparts within
the FWHM of the 15$\mu$m image. All but one of the knots without an
optical counterpart are in the overlap region.

The 15$\mu$m emission from VV 114 is morphologically
very similar to the 3.0 GHz emission, 
with many of
the brightest radio star-forming regions 
having a mid-IR counterpart (Figure \ref{fig:figgray}h). Finally, the CO(2-1) emission shown as contours 
in Figure \ref{fig:figgray}i, which comprises half of the total CO(2-1) emission from
VV114 measured in archival 7$\arcsec$ ALMA data, 
is co-spatial with the VV 114E cores
and extends westward into the overlap region. The highest surface
brightness CO emission is coincide with the optical-obscured knots in the
overlap region.

In order to better interpret the nature of the detected sources, their mid-IR colors were compared against those of the LIRGs in GOALS for which {\it Spitzer} IRS observations are available
\citep{stierwalt2013}.  The IRS spectra were convolved with the MIRI filter transmission functions in order to derive the flux density in each filter. The results are illustrated in  Figure \ref{fig:figplot}. In the plot, the flux density ratio F770W/F560W is effectively a measure of the excess in the F770W filter due to PAH emission and F1500W/F560W is a measure of the mid-IR spectral slope (note that the 6.2$\mu$m PAH feature is redshifted out of F560W for
VV 114). The plot clearly shows a gradient across the F770W/F560W axis as a function of 6.2$\mu$m PAH equivalent width (EQW), with higher values of F770W/F560W corresponding to high PAH EQW. This equivalent width is often used as an AGN diagnostic, with PAH EQW $<0.25$ (red and orange in the plot) being AGN dominant systems \citep[e.g.,][]{armus2007}.

In Figure \ref{fig:figplot}, the mid-IR colors of the mid-IR knots are shown. If the assumption is made that MIRI sources and GOALS LIRGs occupying the same area of the plot have nearly identical mid-IR spectra, the majority of the knots have (1) red dust continuum slopes (F1500W/F560W) and (2) strong PAH emission. The colors are thus consistent with
highly reddened star-forming regions.

The VV 114E cores are also plotted in Figure \ref{fig:figplot}. 
The NE core has star formation-like mid-IR colors, whereas the SW core has a bluer dust continuum and a low PAH equivalent width based on the relatively low F770W/F560W.   

The mid-IR colors of the total VV 114 emission agree well with colors derived from {\it Spitzer} IRS. 
Further, the total MIRI and {\it Spitzer} IRAC measurements of VV 114 ($0.4378\arcmin$-radius aperture) in similar
wavelength bands agree to within 9-20\% -- i.e., F560W ($0.21$ Jy) vs. IRAC 5.8$\mu$m Channel 3 ($0.26$ Jy) and
F770W ($0.80$ Jy) vs. IRAC 8.0$\mu$m Channel 4 ($0.73$ Jy). The flux
densities for the three MIRI bands agree to within 6-30\% of similar ISOCAM filter values reported by \citep[][]{lefloch2002}.

\subsection{Diffuse PAH Emission}

The {\it JWST} MIRI images also show that VV 114 contains 
significant diffuse emission. This emission is most notable at 7.7$\mu$m 
(PAH + dust continuum emission: Figure \ref{fig:figgray}e) --
filamentary structures extend well beyond the mid-IR star-forming
regions (knots). The most prominent
7.7$\mu$m filament extends $\sim 8$ kpc to the south of the nucleus of VV 114E and is cospatial with CO(1-0) emission presented in \citet[]{yun1994} and 
the 0.4$\mu$m dust lane
that obscures the bright mid-IR nucleus from optical view (Figure \ref{fig:figcolor}). In F560W, F770W and F1500W bands, the diffuse component
contributes $\sim$ 50\%, 60\% and 36\% to the total emission from VV 114.
In terms of the mid-IR colors, the diffuse component is blue in F1500W/F560W and has F770W/F560W consistent with strong PAH emission.

\section{Discussion}

The {\it JWST} images in Figures \ref{fig:figcolor} and  \ref{fig:figgray} are clearly illustrative of how the vast improvement in both resolution and sensitivity over space- and ground-based telescopes
has enabled the clearest mid-IR view to date of this dust-obscured galaxy merger. 
The VV 114E nucleus is resolved into two primary cores, with the SW core further resolved into two components.
As summarized in \S \ref{sec:knots}, the mid-IR colors of VV 114E NE are consistent with a reddened star-forming
region. It has been suggested based on the measured high HCN/HCO$^{+}$ 
submillimeter line ratio
of the NE core \citep{iono2013} that it is the location of the
putative AGN hinted at by mid-IR
\citep{lefloch2002} and X-ray \citep{grimes2006} 
observations. The AGN would have to be 
optically-thick at mid-IR wavelengths to be
consistent with the present data. In contrast,
the SW core has colors consistent with 
a low PAH equivalent width source and 
thus may be a more likely AGN candidate. Spatially resolved {\it JWST} mid-IR spectra of VV 114E will be analyzed in a companion paper (Rich et al., in prep).

\begin{figure}[htb!]
    \centering
    \vskip 0.05in
    \includegraphics[scale=0.62]{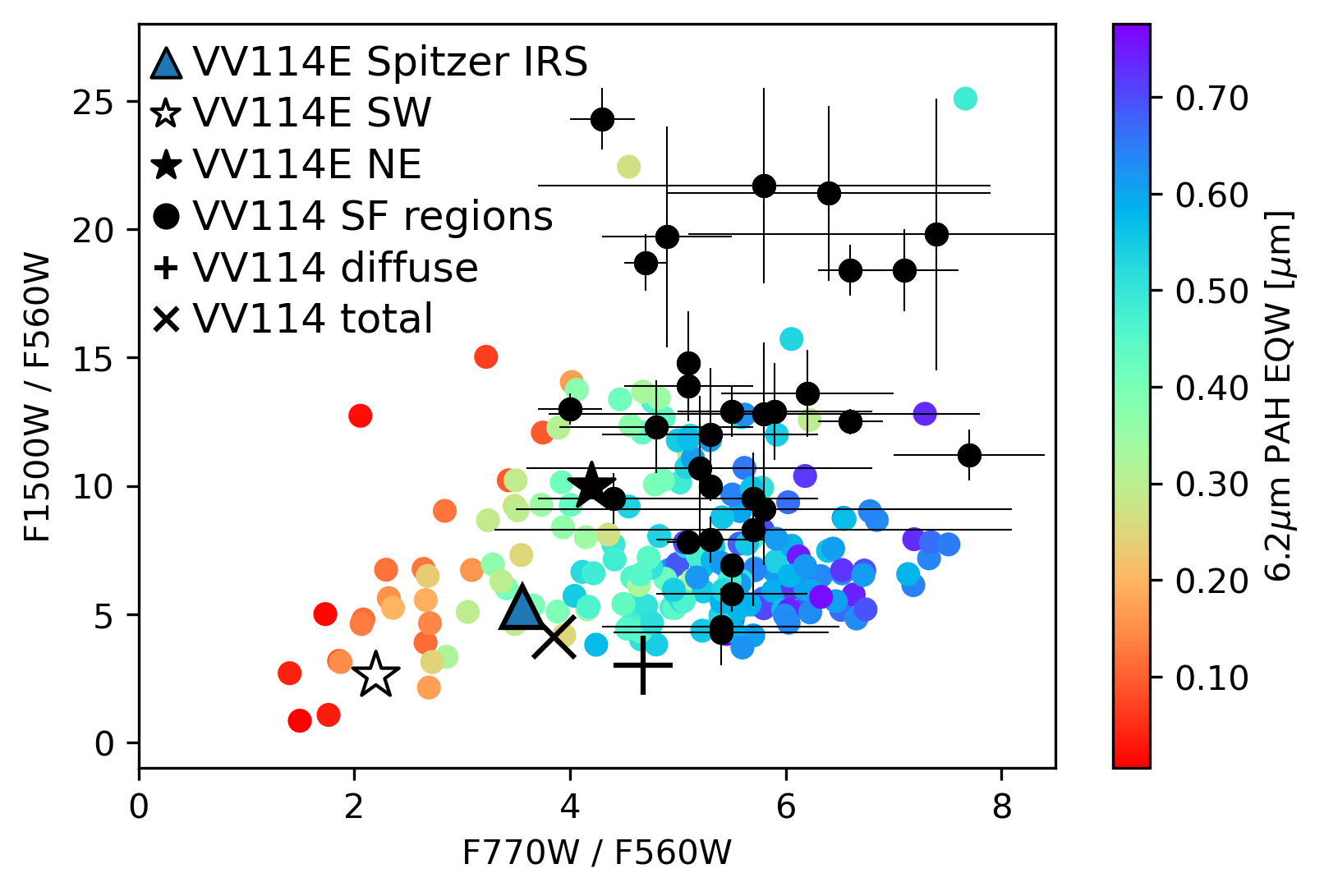}
    \caption{A F770W/F560W vs. F1500W/F560W flux density ratio plot of the VV 114E cores (stars) and the mid-IR knots (filled black circles) in VV 114. The diffuse component of VV 114 is plotted as a ``+'' symbol. Also plotted are LIRGs in
    the GOALS sample with ratios derived from {\it Spitzer} IRS data. The LIRGs are color coded by their 6.2$\mu$m PAH equivalent width, and the triangle is the IRS-derived ratio of VV 114. The MIRI-derived ratios for the total VV 114 emission (the ``$\times$'') agree well with the IRS-derived ratio. The MIRI-detected knots and VV 114E NE have MIRI ratios consistent with LIRGs with strong PAH emission, and they are redder
    in F1500W/F560W relative to the majority of the IRS measured values of the GOALS LIRG sample, indicating that they are reddened star-forming regions. VV 114E SW ratios consistent with an AGN.\label{fig:figplot}}
\end{figure}

The $L_{\rm IR}$ of the nuclear cores can be estimated by adopting the IR 8-1000$\mu$m (W m$^{-2}$)-to-15$\mu$m (W m$^{-2}$ Hz$^{-1}$) luminosity ratio of the GOALS sources that occupy the same area 
of Figure \ref{fig:figplot}. This quantity was measured for each GOALS LIRG by interpolating their {\it IRAS} 12$\mu$m and 25$\mu$m flux densities to
derive the 15$\mu$m flux density and estimating the $L_{\rm IR}$ from {\it IRAS} measurements \citep{sanders2003AJ}. 
The IR/15$\mu$m ratios range from $1.2-2.1\times10^{14}$ Hz,
with a median value of $1.6\times10^{14}$ Hz and a standard deviation of
$0.18\times10^{14}$ Hz.
Applying these ratios yields a IR luminosity
for the NE and
SW cores of $\sim 8\pm0.8\times10^{10}$ L$_\odot$ and $\sim 5\pm0.5\times10^{10}$ L$_\odot$, respectively, which is a combined 30\% of the $L_{\rm IR}$ of VV 114.
The IR and radio luminosities of these cores are roughly consistent with the
IR-to-radio luminosity correlation of star-forming galaxies.
As the sources are unresolved, the resolution of {\it JWST} at 5.6$\mu$m 
($ \theta_{\rm FWHM} \sim 0.18\arcsec = 70$ pc) can be used to derive
a lower limit of the IR luminosity density of
$\Sigma _{\rm IR} \gtrsim 2\pm0.2\times10^{13}$ L$_\odot$ kpc$^{-2}$ and $\gtrsim 6.7\pm0.7\times10^{12}$ L$_\odot$
kpc$^{-2}$, respectively 
(note that the SW core has two components). These numbers are
in the range of that measured for the Orion core 
\citep[$\sim 2\times10^{12}$ L$_\odot$ kpc$^{-2}$:][]{soifer2001} 
in the Galaxy 
and the nuclei of Arp 220 \citep[$\sim 1.5\times10^{13}$ L$_\odot$ kpc$^{-2}$:][]{barcos-munoz2015}.

Approximately 40 knots that are bright at mid-IR wavelengths are identified. These knots have mid-IR colors consistent with star-forming regions (\S \ref{sec:knots}) and are distributed in the overlap region and in VV 114W. At the resolution of the 15$\mu$m image, 75\% of
these knots have optical counterparts.  
These may be the low optical depth 
components of the embedded, mid-IR starbursts, or low optical depth star-forming regions that formed in the vicinity of mid-IR luminous regions. (see Linden et al, in prep. for a discussion of the NIRCAM imaging observations of VV 114.) 
Notably, the knots within the overlap region do not have optical
counterparts.
The knots collectively 
have $L_{\rm IR} \sim 0.02-5\times10^{10}$ L$_\odot$, with 
a median of $9.4\times10^8$ L$_\odot$ and average uncertainty in the IR/15$\mu$m conversion factor of 24\%. By comparison,
the median $L_{\rm IR}$ derived from the IR-radio correlation for VLA-detected star-forming regions
with effective radii $\sim 50-60$ pc (i.e., approximately the resolution of the F560W data) are $6.0\pm4.7\times10^7$ L$_\odot$ and $7.2\pm4.4\times10^9$ L$_\odot$ for normal star-forming galaxies and GOALS LIRGs, respectively \citep[][]{song2022}. 

The most striking feature in the MIRI images is the diffuse emission, especially 
the filamentary structure that covers most of the F770W image. 
This band contains the broad 7.7$\mu$m PAH spectral feature, and it is clear from the image that
PAH is being excited well beyond the regions in VV 114 occupied by bright mid-IR star-forming regions. The filamentary structure is reminiscent of {\it Spitzer} IRAC 8$\mu$m images of nearby spiral galaxies in \citet[]{elmegreen2019};
the authors attribute the high $8.0\mu{\rm m} / 5.8\mu{\rm m}$ to
strong PAH emission in the 8.0$\mu$m filter. If the diffuse component
in the F560W has the same flux density as the dust continuum 
in the diffuse component in the F770W image, then 
$\sim 80\%$ of the diffuse emission in F770W image is PAH emission.

These new {\it JWST} observations of VV 114 allow for the first time to measure the properties of the most obscured IR-luminous star-forming regions and 
to quantify the contributions of diffuse emission to mid-IR light in the extreme environment of a luminous galaxy merger.


\section{Summary}

New {\it JWST} MIRI observations of VV 114 are presented. The following conclusions are
reached:

\vskip 0.1in

\noindent
(1) The imbedded eastern nucleus of VV 114 is resolved into a NE and SW core. The mid-IR colors of these cores are consistent with a starburst
(NE) and an AGN (SW).
The lower limits of their estimated 
$\Sigma _{\rm IR}$ lie between the Orion core and the nuclei of Arp 220.

\vskip 0.1in

\noindent
(2) Approximately 40 star-forming regions (knots) bright at mid-IR wavelengths
are identified, 25\% of which have no optical counterpart. All but one of the knots with no optical counterpart lie in the overlap region. The knots have
$L_{\rm IR} \sim 0.02-5\times10^{10}$ L$_\odot$. These values are above the
median $L_{\rm IR}$ of comparably-size radio-identified star-forming regions in nearby, normal star-forming galaxies,
but overlap the median $L_{\rm IR}$ of radio-identified GOALS LIRGs.

\vskip 0.1in

\noindent
(3) Diffuse emission accounts for a significant fraction of the light in
each MIRI image. In particular, filamentary emission is detected at 7.7$\mu$m that accounts for 60\% of the light at this wavelength.
Approximately 80\% of this extended component is estimated to be PAH emission heated by the star-forming regions and
underlying stellar population.


\vskip 0.2in
\noindent
The research was supported by NASA grants JWST-ERS-01328 and HST-GO15472. 
Y.S. was funded in part by the NSF through the Grote Reber Fellowship Program administered by Associated Universities, Inc./National Radio Astronomy Observatory. 
VU acknowledges funding support from NASA Astrophysics Data Analysis Program (ADAP) grant 80NSSC20K0450. 
S.A. gratefully acknowledges support from an ERC Advanced
Grant 789410, from the Swedish Research Council and from the Knut and Alice Wallenberg (KAW) Foundation.
The National Radio Astronomy Observatory is a facility of the National Science Foundation operated under cooperative agreement by Associated Universities, Inc. The authors acknowledge the usage of the NASA/IPAC Infrared Science Archive, which is funded by the National Aeronautics and Space Administration and operated by the California Institute of Technology, and CARTA \citep[][]{comrie2021zndo}. 

\facilities{JWST, HST, Spitzer, ALMA, MAST, IRSA, NASA/ADS, IPAC/NED}


\begin{deluxetable*}{rrrrrr}
\tabletypesize{\footnotesize}
\tablecolumns{5}
\tablewidth{0pt}
\tablenum{1}
\tablecaption{MIRI Photometry of VV 114E Cores and Compact Star-forming Knots \label{tab:phot}}
\tablehead{ \colhead{ID} &
\colhead{RA (J2000.0)} & \colhead{Dec (J2000.0)} & \colhead{F560W} & \colhead{F770W} & \colhead{F1500W} \\
\colhead{} & \colhead{(deg)} & \colhead{(deg)} & \colhead{(mJy)} & \colhead{(mJy)} & \colhead{(mJy)}}
\startdata
(NE) 1 & 16.948352 &	-17.506764 &	$19\pm7$ & $90\pm2$ &	$190\pm6$ \\
(SW) 2 & 16.947965 &	-17.507016 &	$73\pm10$ & $160\pm8$ &	$190\pm6$ \\
\hline
3	&	16.942842	&	-17.507251	& $	0.17	\pm	0.001	        $ & $	0.85	\pm	0.01	         $ & $	2.5	\pm	0.3	$	\\
4	&	16.942898	&	-17.507999	& $	0.24	\pm	0.006	        $ & $	1.1	\pm	0.05	         $ & $	4.5	\pm	0.2	$	\\
5	&	16.942909	&	-17.507004	& $	0.045	\pm	0.005	$ & $	0.22	\pm	0.01	         $ & $	0.88	\pm	0.16	$	\\
6	&	16.9433	    &	-17.508007	        & $	0.066	\pm	0.002	$ & $ <	0.29	                 $ & $	1.1	\pm	0.2	$	\\
7	&	16.9434	    &	-17.508212.        	& $	0.14	\pm	0.02	                 $ & $	0.82	\pm	0.04	$ & $	1.8	\pm	0.1	$	\\
8	&	16.943513	&	-17.507626	& $ <	0.051	                $ & $ <	0.71	                 $ & $	0.49	\pm	0.02	$	\\
9	&	16.943553	&	-17.507837	& $ <	0.044	                 $ & $ <	0.25	                 $ & $	1.1	\pm	0.1	$	\\
10	&	16.943728	&	-17.508898	& $	0.084	\pm	0.03    	$ & $	0.48	\pm	0.13	         $ & $	0.70	\pm	0.13	$	\\
11	&	16.944169	&	-17.506207	& $	0.34	\pm	0.02	                 $ & $	2.4	\pm	0.05	$ & $	6.2	\pm	0.3	$	\\
12	&	16.944281	&	-17.507248	& $	0.46	\pm	0.007	         $ & $	1.8	\pm	0.1	$ & $	6.0	\pm	0.3	$	\\
13	&	16.944435	&	-17.506666	& $ <	0.051	                 $ & $ <	0.16	                 $ & $	0.43	\pm	0.12	$	\\
14	&	16.944479	&	-17.506386	& $	0.10	\pm	0.05	                 $ & $	0.61	\pm	0.30	$ & $	1.0	\pm	0.3	$	\\
15	&	16.944516	&	-17.505267	& $	0.12	\pm	0.006	        $ & $	0.63	\pm	0.04	        $ & $	0.93	\pm	0.10	$	\\
16	&	16.944565	&	-17.505692	& $	0.62 \pm	0.13	                 $ & $	3.6	\pm	1.2	$ & $	5.6	\pm	0.6	$	\\
17	&	16.944635	&	-17.507562	& $	1.0	\pm	0.08	                 $ & $	5.3	\pm	0.4	$ & $	14	\pm	1.0	$	\\
18	&	16.944663	&	-17.50781	        & $	0.73	\pm	0.04.            	$ & $	3.1	\pm	0.2	         $ & $	18	\pm	0.2	$	\\
19	&	16.944907	&	-17.508914	& $	0.18	\pm	0.01	                 $ & $	1.0	\pm	0.03	$ & $	2.3	\pm	0.1	$	\\
20	&	16.944914	&	-17.507575	& $	0.63	\pm	0.14.       	        $ & $	3.6	\pm	1.0	        $ & $	8.0	\pm	0.3	$	\\
21	&	16.944943	&	-17.506685	& $	0.22	\pm	0.02	                $ & $	1.7	\pm	0.1	        $ & $	2.5	\pm	0.1	$	\\
22	&	16.944956	&	-17.506519	& $	0.270	\pm	0.03	       $ & $	1.7	\pm	0.1	        $ & $	3.7	\pm	0.3	$	\\
23	&	16.945271	&	-17.507432	& $	0.12	\pm	0.02.           	$ & $	0.70	\pm	0.22	        $ & $	2.6	\pm	0.2	$	\\
24	&	16.945378	&	-17.508825	& $	0.15	\pm	0.01	               $ & $	0.80	\pm	0.07	        $ & $	0.8	\pm	0.1	$	\\
25	&	16.945624	&	-17.507555	& $	0.20	\pm	0.04	               $ & $ <	1.1	                 $ & $	1.4	\pm	0.1	$	\\
26	&	16.945726	&	-17.508668	& $	0.14	\pm	0.001	      $ & $	0.74	\pm	0.007	$ & $	0.93	\pm	0.04	$	\\
27	&	16.945768	&	-17.507342	& $	0.55	\pm	0.02	              $ & $	2.8	\pm	0.04	        $ & $	4.3	\pm	0.1	$	\\
28	&	16.946031	&	-17.508483	& $	0.037	\pm	0.005	$ & $	0.24	\pm	0.05	        $ & $	0.80	\pm	0.05	$	\\
29	&	16.946061	&	-17.506594	& $	0.20	\pm	0.05	              $ & $	1.4	\pm	0.2	        $ & $	3.9	\pm	0.1	$	\\
30	&	16.946127	&	-17.507399	& $	0.77	\pm	0.12	               $ & $	4.1	\pm	0.4	        $ & $	9.2	\pm	1.4	$	\\
31	&	16.946162	&	-17.508302	& $	0.036	\pm	0.01	       $ & $ <	0.44	                 $ & $	1.2	\pm	0.1	$	\\
32	&	16.946188	&	-17.506119	& $	0.094	\pm	0.001	$ & $	0.62	\pm	0.03	        $ & $	1.7	\pm	0.1	$	\\
33	&	16.946347	&	-17.506782	& $	0.53	\pm	0.05	                 $ & $	2.4	\pm	0.3	$ & $	5.0	\pm	0.3	$	\\
34	&	16.946579	&	-17.504611	& $	0.066	\pm	0.006	$ & $	0.38	\pm	0.02	        $ & $	0.63	\pm	0.04	$	\\
35	&	16.946696	&	-17.50709	         & $	0.48	\pm	0.01	                 $ & $	2.6	\pm	0.5   $ & $	2.0	\pm	0.3	$	\\
36	&	16.946775	&	-17.506319	& $	0.38	\pm	0.02	                $ & $	2.5	\pm	0.04	        $ & $	4.8	\pm	0.02	$	\\
37	&	16.947041	&	-17.507299	& $	1.09	\pm	0.1	                $ & $	5.2	\pm	0.8	        $ & $	13	\pm	1	$	\\
38	&	16.947402	&	-17.506981	&  \nodata	                        & $	4.5	\pm	0.1	        $ & $	6.1	\pm	0.7	$	\\
39	&	16.947425	&	-17.507211	& $	1.8	\pm	0.4	                $ & $	9.4	\pm	1.9	        $ & $	19	\pm	2	$	\\
40	&	16.948176	&	-17.511708	& $	0.047	\pm	0.0004	$ & $	0.25	\pm	0.004	$ & $	0.47	\pm	0.02	$	\\
\enddata
\end{deluxetable*}

\bibliography{reference.bib}

\end{document}